\documentclass[preprint,12pt]{elsarticle}

\usepackage{amsmath, amsfonts, amssymb}
\usepackage[margin=0.75in]{geometry}
\usepackage{graphicx}
\usepackage{hyperref}
\usepackage{setspace}
\usepackage{algorithm}
\usepackage{algpseudocode}
\usepackage{tabularx}
\usepackage{lscape} 
\usepackage{rotating}
\usepackage{hyphenat} 
\usepackage{enumitem}
\usepackage{bm}
\usepackage{placeins}
\usepackage{setspace}
\usepackage[capitalise]{cleveref}
\usepackage{fancyhdr}
\usepackage{color}
\usepackage{lineno}
\usepackage{adjustbox} 
\usepackage{tikz} 
\usepackage{caption}

\usepackage{cite}

\usepackage{soul}

\usepackage{array, booktabs, makecell} 

\def \eg{e.g.\ }

\newcommand{\edit}[1]{\color{black}#1 \color{black}}

\begin{document}

\begin{frontmatter}
\title{Physics-constrained coupled neural differential equations for one dimensional blood flow modeling}
\author[utah1,utah2]{Hunor Csala}
\author[lanl]{Arvind Mohan}
\author[lanl]{Daniel Livescu}
\author[utah1,utah2]{Amirhossein Arzani\corref{cor1}}
\ead{amir.arzani@sci.utah.edu}

\address[utah1]{Department of Mechanical Engineering, University of Utah, Salt Lake City, UT, USA}
\address[utah2]{Scientific Computing and Imaging Institute, University of Utah, Salt Lake City, UT, USA}
\address[lanl]{Computational Physics and Methods, Los Alamos National Laboratory, Los Alamos, NM, USA}
\cortext[cor1]{Corresponding author}

\begin{abstract}

\textbf{Background:} Computational cardiovascular flow modeling plays a crucial role in understanding blood flow dynamics. While 3D models provide acute details, they are computationally expensive, especially with fluid-structure interaction (FSI) simulations. 1D models offer a computationally efficient alternative, by simplifying the 3D Navier-Stokes equations through axisymmetric flow assumption and cross-sectional averaging. However, traditional 1D models based on finite element methods (FEM) often lack accuracy compared to 3D averaged solutions. 

\textbf{Methods:} This study introduces a novel physics-constrained machine learning technique that enhances the accuracy of 1D cardiovascular flow models while maintaining computational efficiency. Our approach, utilizing a physics-constrained coupled neural differential equation (PCNDE) framework, demonstrates superior performance compared to conventional FEM-based 1D models across a wide range of inlet boundary condition waveforms and stenosis blockage ratios. A key innovation lies in the spatial formulation of the momentum conservation equation, departing from the traditional temporal approach and capitalizing on the inherent temporal periodicity of blood flow. 

\textbf{Results:} This spatial neural differential equation formulation switches space and time and overcomes issues related to coupling stability and smoothness, while simplifying boundary condition implementation. The model accurately captures flow rate, area, and pressure variations for unseen waveforms and geometries\edit{, having 3-5 times smaller error than 1D FEM, and less than 1.2\% relative error compared to 3D averaged training data.} We evaluate the model's robustness to input noise and explore the loss landscapes associated with the inclusion of different physics terms. 

\textbf{Conclusion:} This advanced 1D modeling technique offers promising potential for rapid cardiovascular simulations, achieving computational efficiency and accuracy. By combining the strengths of physics-based and data-driven modeling, this approach enables fast and accurate cardiovascular simulations. 
\end{abstract}


\end{frontmatter}

\pagebreak

\section{Introduction}
\label{sec:intro}

Cardiovascular diseases are a major cause of mortality worldwide, emphasizing the need to improve our understanding of circulatory physiology and related disorders. Cardiovascular flow modeling has emerged as a powerful tool in this endeavor, offering insights into complex hemodynamics, aiding in the development of medical devices, and assisting with clinical decision-making~\citep{schwarz2023beyond}. These models range from simplified lumped parameter representations to highly detailed three-dimensional fluid-structure interaction (FSI) simulations. While the latter provide intricate detail, they often require substantial computational resources, limiting their practical applicability in clinical scenarios. This challenge has spurred the development of reduced order models (ROMs), which aim to capture the essential dynamics of the cardiovascular system while significantly reducing computational complexity. In the field of cardiovascular fluid mechanics, two main categories of ROMs have emerged: physics-based and data-driven approaches~\citep{arzani2021data,pfaller2024reduced,macraild2024accelerated}. Physics-based ROMs leverage fundamental principles of fluid dynamics and vascular mechanics to create simplified yet physically meaningful and interpretable representations. These models often rely on assumptions such as axisymmetric flow or linearized equations to achieve computational efficiency. On the other hand, data-driven ROMs utilize advanced statistical and machine learning techniques to extract low-dimensional representations from high-fidelity simulation data or experimental measurements.
Physics-based ROMs using spatial dimensionality reduction have been popular in the cardiovascular research community. As an example, 0D Windkessel models have been widely used and are recognized by the clinical community~\citep{westerhof2009arterial}. Both 1D and 0D models have been shown to produce fast estimates of blood flow variables~\citep{pfaller2022automated, grinberg2011modeling}. 1D models use a cross-sectionally averaged formulation of the Navier-Stokes equations, where the single velocity component points along the centerline direction. 0D models solve a system of ordinary differential equations (ODEs), building on electrical circuit analogy, defining circuit elements based on the geometric and material properties of the vessel segments.

While 1D models have demonstrated generally satisfactory performance, they still face challenges in accurately representing specific scenarios. These limitations become particularly evident in modeling flow patterns in complex geometries such as aneurysms or stenoses, and precisely characterizing pressure and flow variations across bifurcations~\citep{pfaller2022automated,grinberg2011modeling, xiao2014systematic, grande20221d,rubio2024hybrid, reymond2012patient}. Without special treatment, the 1D models are known to underestimate the pressure drop along stenosed arteries. Correction formulas based on empirical relations have been proposed~\citep{seeley1976effect, young1973flow} and used together with the 1D models~\citep{pfaller2022automated, grande20221d, blanco2018comparison}. These empirical formulas relate pressure drop to various factors such as flow rate, cross-sectional areas, stenosis length, and material properties, providing a more accurate representation of fluid dynamics in complex regions.


Traditional modeling approaches have made significant strides in addressing the complexities of cardiovascular flows, but some challenges persist, and there is room for further improvement in existing solutions. In recent years, machine learning techniques have emerged as complementary tools, offering new perspectives on both longstanding and novel challenges in cardiovascular flow problems~\citep{macraild2024accelerated,arzani2022machine, arzani2021data}. These include data assimilation~\citep{habibi2021integrating}, denoising~\citep{csala2024comparison}, super-resolution~\citep{fathi2020super}, reduced order modeling~\citep{pegolotti2024learning}, segmentation~\citep{kong2024sdf4chd}, phase error correction~\citep{you2022deep}, treatment planning~\citep{tanade2024harvi}, among others. Some commonly used techniques include proper orthogonal decomposition (POD)~\citep{zainib2021reduced, siena2023data}, dynamic mode decomposition (DMD)~\citep{habibi2021integrating}, physics-informed neural networks (PINNs)~\citep{arzani2021uncovering}, graph neural networks (GNNs)~\citep{pegolotti2024learning}, convolutional autoencoders (CAEs)~\citep{gharleghi2022transient} and generative adversarial networks (GANs)~\citep{kong2024sdf4chd}.

While machine learning is being widely applied to 3D cardiovascular flow models, 0D and 1D-based blood flow models have received comparatively less attention. This is primarily because these lower-dimensional models are already computationally efficient compared to their 3D counterparts. However, machine learning has the potential to improve their accuracy, which is lost due to their geometric dimensionality reduction.  Recent research has begun to explore how machine learning can enhance 0D and 1D models, particularly in addressing limitations such as missing boundary conditions or unaccounted physics.~\citet{sen2024PIGNN} used a graph-based PINN model to account for missing boundary conditions in the 1D framework by using experimental data.~\citet{grigorian2024hybrid} used a neural ODE-based framework for capturing ventricular interactions in a hybrid 0D model with synthetic data. ~\citet{li2021oneDvocal} applied the sparse identification of nonlinear dynamics (SINDy) method to discover unknown terms in the 1D equations of vocal fold vibrations, sharing many similarities with 1D blood flow equations. These studies demonstrate the potential of machine learning to further improve the accuracy and applicability of physics-based reduced-order cardiovascular models.

Neural Differential Equations (NDEs)~\citep{chen2018NODE, rackauckas2020universal} represent a powerful intersection of machine learning and scientific computing, connecting neural networks (NNs) and differential equations. This approach leverages the insight that residual connections in neural networks can be interpreted as Euler discretizations of ODEs. By extending this concept, NDEs allow for the incorporation of neural networks directly into differential equation systems, enabling the capture of complex, unknown physics or the learning of dynamics from data. There is growing interest in extending these approaches to partial differential equations (PDEs). They can be used to augment existing PDE models with learned components, addressing scenarios where the underlying physics is partially known but certain terms or parameters remain elusive~\citep{ramadhan2020capturing}. For instance, NDEs can help model subgrid-scale effects in turbulence simulations~\citep{sanderse2024scientific, kang2023learning} or capture complex constitutive relationships in solid mechanics~\citep{masi2024neural}. These methods, also known as differentiable programming or differentiable simulators~\citep{holl2024phiflow, bezgin2024jax}, integrate gradient-based optimization with traditional scientific computing, enabling end-to-end training of hybrid models. A key challenge is the need for differentiable differential equation solvers that allow backpropagation of gradients through the computational graph, which is essential for effective training using gradient-based optimization techniques.

Neural PDE approaches have been widely applied to various physical systems, often focusing on PDEs with periodic spatial boundary conditions. Examples include homogeneous isotropic turbulence~\citep{shankar2022validation}, Kuramoto-Sivashinsky equation~\citep{gelbrecht2021NPDE}, Burgers' equation~\citep{melchers2023_neuralPDE}, and Fisher-KPP equation~\citep{rackauckas2020universal}. In these approaches, space is typically discretized and a neural ODE problem is formulated for the time evolution of the system. However, applying these techniques to cardiovascular flows presents unique challenges. Blood flow dynamics are characterized by temporal periodicity due to the rhythmic nature of the heartbeat. Moreover, the inlet boundary condition is usually a time-dependent pulsatile flow rate or pressure waveform. These characteristics make it difficult to directly apply traditional neural PDE methods to cardiovascular systems.

To address these challenges, we propose a novel approach of switching the roles of space and time in our neural PDE framework. The idea is to discretize time and formulate the differential equations in space. This innovative technique capitalizes on the inherent temporal periodicity of blood flows. Unlike most fluid dynamics applications of machine learning, cardiovascular flows do not require solutions for arbitrary future times. Instead, it is sufficient to obtain solutions at a finite number of time instances over a cardiac cycle due to the temporal periodicity. This unique scenario creates an opportunity where switching space and time could be particularly beneficial, potentially leading to improved stability and accuracy. The concept of exploiting temporal periodicity in cardiovascular flow modeling has been recently utilized by~\citet{esmaily2023stabilized}. They developed a time-spectral formulation using Fourier transforms in time, rather than the traditional space approach in spectral methods~\citep{boyd2001chebyshev}, and applied it to 3D finite element method (FEM) based models of cardiovascular flows. Their approach focuses on 3D FEM modeling in the frequency domain and it was shown to speed up blood flow simulations by orders of magnitude compared to the traditional formulation.


The main goal here is to bridge the gap between 1D and 3D models by creating a physics-based data-driven model that is still 1D in essence but achieves greater accuracy than conventional 1D approaches since it is trained with 3D averaged data. A novel physics-constrained coupled differential equation (PCNDE) framework is introduced that leverages 3D averaged data for training, but also includes some of the physics from previous models, like the continuity equation and the pressure drop formula, with correction terms to account for missing physics or deficient assumptions. Two key novel ideas are introduced. First, switching space and time in the NDE framework simplifies the treatment of boundary conditions and helps the PDE coupling stability and smoothness. Second, a Fourier series is exploited for hard constraining temporal periodicity of the cross-sectional area deformation model. The new method is demonstrated on different idealized stenosis geometries and a wide range of inlet boundary condition flow rate waveforms, and compared with the state-of-the-art 1D FEM solver. \edit{Our approach strikes a balance between traditional physics-based 1D models and ROMs trained on 3D data by leveraging 3D-derived data for enhanced accuracy while maintaining the computational efficiency and robustness of a 1D framework. This hybrid method ensures faster inference while addressing the challenges associated with purely data-driven ROMs and the lack of accuracy of traditional 1D models.}\edit{Modal decomposition-based surrogate models~\citep{habibi2021integrating,siena2023data} are a significant related area of research, enabling full 3D reconstructions. These models can be either physics-based, such as those using Galerkin projection or entirely data-driven. While traditionally constrained by linearity, recent advancements have focused on incorporating nonlinearity through neural networks~\citep{siena2023data}. In contrast, our PCNDE framework provides a 1D focused alternative with built-in physics constraints, offering robust performance and efficiency. This positions PCNDE as a complementary approach, suited for different modeling scenarios where computational simplicity, noise resilience, and adaptability are priorities.} The present study has the following novel contributions:
\begin{itemize}
  \item Switching space and time in the NDE framework is proposed, leading to improved smoothness and stability. The temporal periodicity of blood flow is the main motivation behind this shift of perspective.
  \item Handling coupled PDEs with time-dependent boundary conditions is demonstrated in the NDE framework. Training stability is discussed in relation to different physics terms.
  \item The temporal periodicity of area predictions is embedded as an inductive bias in the model through the use of Fourier series.
  \item While the machine learning model is essentially 1D in formulation, it achieves higher accuracy than traditional 1D blood flow equations by incorporating 3D averaged data for training.
  \item Two-way coupled FSI simulations are used for training the model taking into account the arterial wall movement.

\end{itemize}

The rest of the paper is structured in the following way. In Section~\ref{sec:methods}, the 1D and 3D models are described for generating data and the PCNDE framework is introduced. In Section~\ref{sec:results}, the PCNDE results are presented, analyzed, and compared to the 1D FEM solution. In Section~\ref{sec:discussion}, the results are discussed and strengths and limitations are described. In Section~\ref{sec:conclusion}, the final conclusions are drawn.

\section{Methods}

Below, the 1D and 3D blood flow approaches are explained. The 1D FEM solution will serve as a benchmark, while the 3D FSI model will be used to generate training data. Next, the PCNDE approach is introduced and the data generation process and model setup are presented.

\label{sec:methods} 
\subsection{1D Blood Flow Equations}
By cross-sectional averaging the 3D Navier-Stokes equations along a vessel centerline, the 1D equations of blood flow can be formulated~\citep{hughes1973one, hughes1974_1d}. For Newtonian incompressible flow with deformable walls, the continuity and momentum equations are the following:
\begin{align}
    \label{eq:cont}
    \dfrac{\partial S}{\partial t} &= - \dfrac{\partial Q}{\partial z} \;,\\
    \label{eq:momentum}
    \dfrac{\partial Q}{\partial t} &= - (1+\delta) \dfrac{\partial}{\partial z}\left( \dfrac{Q^2}{S} \right) - \dfrac{S}{\rho} \dfrac{\partial p}{\partial z} +  N \dfrac{Q}{S} + \nu \dfrac{\partial^2 Q}{\partial z^2} \;,
\end{align}
where \(Q\) is flow rate, \(S\) is cross-sectional area, \(p\) is pressure, \(t\) is time, \(z\) is the axial coordinate along the centerline, \(\rho\) is blood density, and \(\nu\) is kinematic viscosity. \(\delta\) and \(N\) are two parameters related to the velocity profile assumption. In the case of a parabolic profile, the following relations hold~\citep{hughes1973one}:
\begin{align}
    \delta &= \dfrac{1}{3} \;,\\
    N &= - 8 \pi \nu \;.
\end{align}

Usually, a pulsatile flow rate waveform is specified as the inlet boundary condition:
\begin{align}
Q(z=0,t) = Q_{in}(t) \;.
\end{align}
\(Q_{in}(t)\) could be either a patient-specific waveform obtained from experimental measurements or a population-averaged waveform for the given artery. The outlet boundary conditions are usually 0D Windkessel models. The initial condition is \(Q(z,t=0) = 0\), and since the inlet boundary condition is periodic in time, the system will converge to the periodic solution. 

An additional closure equation is needed to close the system of equations. A constitutive equation is used that relates the pressure to the cross-sectional area and the wall's material properties. The most commonly used models are the linear and Olufsen models~\citep{pfaller2022automated,olufsen1999structured}:
\begin{align}
    p(z,t) &= p_{ref} + \dfrac{4}{3} \dfrac{Eh}{r_{0}}\left(\sqrt{\dfrac{S(z,t)}{S_{u}(z)}} -1 \right) \;, \\
    p(z,t) &= p_{ref} + \dfrac{4}{3}\left( k_{1} e^{k_{2} r_{0}(z)} + k_{3} \right) \left( 1 - \sqrt{\dfrac{S_{u}(z)}{S(z,t)}} \right) \;,
\end{align}
where \(k_{1}, k_{2}, k_{3}\) are empirically fitted coefficients, \(p_{ref}\) is a reference pressure, \(S_{u}(z)\) is the undeformed area, \(E\) is the Young modulus of the wall, \(h\) is the wall thickness, and \(r_{0}\) is a reference radius. In the present study, the linear model will be used for the 1D FEM simulations. 

\subsection{3D Blood Flow Model}
The 3D incompressible Navier-Stokes equations are used to describe the flow of blood in the cardiovascular system. The SimVascular software package~\citep{updegrove2017simvascular}, an open-source finite-element computational fluid dynamics (CFD) solver, was used for the 3D simulations. A Newtonian blood viscosity model was used with the density of \(\rho = 1060 \) kg/m\(^{3}\) and dynamic viscosity of \(\mu = 0.004\) kg/ms. The coupled momentum method~\citep{figueroa2006coupled} was used for the wall movement, which is a two-way coupled fluid-structure interaction (FSI) technique. The vessel wall movement is described using a linear membrane model. The wall deformations affect the fluid through a nonzero normal wall velocity, while the solid-fluid interface mesh is the same as the fluid mesh at the wall. Therefore, the computational time is considerably lower than that of more detailed FSI techniques. For further details, readers are referred to~\citep{figueroa2006coupled}. The other details of the 3D simulations will be described along the data generation workflow in Sec.~\ref{sec:data}.

\subsection{Temporal Neural PDEs}
\label{sec:temporal_NODE}
The general neural differential equation approach is to parameterize parts of the right-hand side of Eq.~\ref{eq:momentum} with neural networks
\begin{align}
    \dfrac{\partial Q}{\partial t} = f_{\theta}(Q) + h(Q,S,z,t) \;,
\end{align}
where \(f_{\theta}\) is a neural network with \(\theta\) representing the optimizable weights and biases, while \(h\) is a function that could represent physics terms kept from Eq.~\ref{eq:momentum}. In this formulation, space is discretized into \(n_{z}\) spatial coordinates and one ODE is solved for each spatial point throughout time. A self-implemented differentiable RK4 ODE solver was used for this case, with the inlet boundary condition being enforced after each timestep. Importantly, we ensured that the boundary condition enforcement procedure itself is differentiable, a non-trivial task that required specific implementation techniques.  An illustratory example of the temporal neural ODE formulation was done, using a model where all the right-hand side of the momentum equation is parameterized with a NN, i.e., \(h(Q,S,z,t)=0\). \edit{A second example will be discussed, where time is also input to the neural network, thus the right-hand side has the form of \(f_{\theta}(Q,t)\).} The models were trained with a mean squared error (MSE) loss between the solution of the differential equations and the ground-truth FSI data. After obtaining the flow rate \(Q\), the 1D continuity equation (Eq.~\ref{eq:cont}) was solved for the area. 

The temporal formulation exhibits challenges in maintaining spatial smoothness of the solution, resulting in inaccurate spatial derivatives and, consequently, erroneous area predictions as shown later in Sec.~\ref{sec:temporal_NPDE_results}. This motivated the development of the spatial neural PDE approach presented in the next section. We presented the proof-of-concept results for the space-time switching previously in~\citep{csala2023modeling}, utilizing 1D training data only. It was shown that the model can generalize well for a wide range of unseen inlet boundary conditions. 


\subsection{Physics-constrained Coupled Neural Differential Equations (PCNDE)}
We introduce a novel approach by reformulating the momentum equation as a neural PDE in space rather than time. This spatial formulation effectively addresses the smoothness and coupling issues inherent in the temporal formulations. The full PCNDE framework involves a multi-step fitting process. First, a neural PDE problem is formulated for the momentum equation and solved in space to obtain the flow rate. Next, a two-stage model is used for predicting the cross sectional area. The first stage is built upon the 1D continuity equation with additional learnable correction terms. The second stage involves a Fourier series fit that snaps the area predictions to be explicitly periodic in time. The momentum and continuity equations are two-way coupled in an iterative manner. After the flow rate and area models are trained, another two stage fitting process is used for modeling the pressure variations. The first stage utilizes a small neural network to capture the temporal variations. Finally, the second pressure stage uses an analytical function to model the spatial variations of pressure. This algorithm is tested on idealized stenosis geometries, where vessel narrowing due to blockage creates flow separation and recirculation zones, presenting challenges for traditional 1D models. Detailed descriptions of the PCNDE process, test cases, and data generation are provided in the following sections.

\subsubsection{Flow Rate and Area}

\begin{figure}[h]
    \centering
    \includegraphics[width=0.8\linewidth]{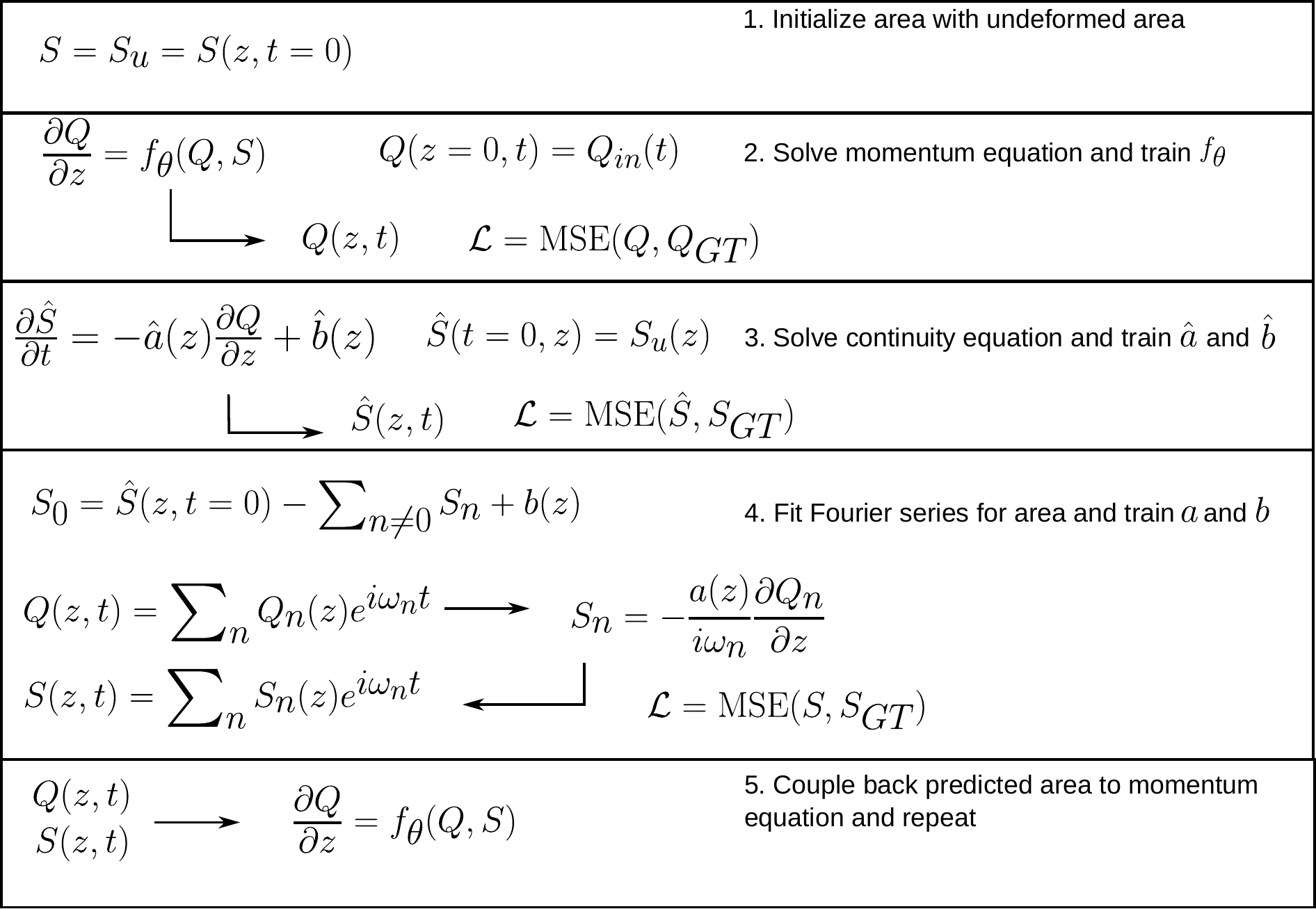}
    \caption{An overview of the Physics-constrained Coupled Neural Differential Equations (PCNDE) algorithm is presented. A spatial neural PDE-based momentum equation is used for predicting the flow rate values, then an initial area approximation is formulated based on the 1D continuity equation with two correction terms. A Fourier series is fit for the final area prediction, enforcing explicit temporal periodicity. The area prediction is coupled back to the momentum equation, and the process is iterated until convergence.}
    \label{fig:methods}
\end{figure}

To overcome the previously mentioned limitations of the temporal formulation, the neural differential equations for the momentum equation have been rewritten in space, and the continuity equation has been  formulated as:
\begin{align}
    \label{eq:PCNDE_mom}
    \dfrac{\partial Q}{\partial z} &= f_{\theta}(Q,S) \;,\\
    \dfrac{\partial \hat{S}}{\partial t} &= -\hat{a}(z) \dfrac{\partial Q}{\partial z} + \hat{b}(z) \;,
    \label{eq:PCNDE_cont}
\end{align}
where \(\hat{a}(z)\) and \(\hat{b}(z)\) are two vectors with optimizable parameters.  The initial condition for the spatial momentum equation is the inlet boundary condition \(Q(z=0,t)=Q_{in}(t)\), and the differential equations are solved in space on the domain \(z\in [0, L]\). That is, in the spatial neural PDE formulation, the original inlet boundary condition becomes an initial condition, facilitating the implementation. The cardiac cycle is discretized into \(n_{t}=100\) time instances, and the solution is evolved in space for all time instances. \(f_{\theta}\) is a fully connected NN with 6 layers of size \(2n_{t}, 8, 8, 8, 8, n_{t}\), and \(\tanh\) activation functions in between. The inputs to the NN \(f_{\theta}\) are flow rate \(Q\) and area \(S\) values concatenated at \(n_{t}\) time points. Therefore, the input layer is of size \(2 n_{t}\). The output of the NN is of size \(n_{t}\), matching the temporal size of \(Q\). An MSE loss was used between the predicted flow rate \(Q\) and the ground-truth values \(Q_{GT}\) to train the NN. The loss was evaluated at all \(n_{z}\) spatial points where data was available.

For the continuity equation, the area is initialized with the undeformed area \(\hat{S}(z,t=0)=S_{u}(z)\). The continuity equation is solved in time for three cardiac cycles \(t \in [0, 3 T]\) to get rid of the initial transients due to the initialization with the undeformed area. Space is discretized into \(n_{z}=101\) points along the centerline. The original 1D continuity equation, Eq.~\ref{eq:cont}, is used as a starting point, enhanced with two correction parameters \(\hat{a}\) and \(\hat{b}\), which are trainable. These corrections are needed since the ground-truth data comes from 3D averaged simulations and does not satisfy the 1D equations. These correction parameters \(\hat{a}\) and \(\hat{b}\) are trained with an MSE loss between the predicted area \(\hat{S}\) and ground-truth area \(S_{GT}\) from the 3D averaged data. \(\hat{a}(z)\) and \(\hat{b}(z)\) vary in space, therefore are of size \(n_{z}\). The spatial derivative \(\partial /\partial z\) in the continuity equation is taken using a second-order central difference scheme.

The total number of optimizable parameters in \(f_{\theta}\) is 2724. The coupled system of PDEs is solved sequentially. First, Eq.~\ref{eq:PCNDE_mom} is solved, and the neural network \(f_{\theta}\) is trained with the undeformed area \(S = S_{u}\) being the second input. Then, the obtained flow rate \(Q\) is used in Eq.~\ref{eq:PCNDE_cont} to solve for an intermediary \(\hat{S}\) and train \(\hat{a}\) and \(\hat{b}\). 

The temporal periodicity of the area predictions is enforced using an additional step involving fitting a Fourier series in time on the area predictions. The cross-sectional area \(S\) and flow rate \(Q\) can be expressed as following using Fourier-series
\begin{align}
    S(z,t) &= \sum_{n} S_n(z) e^{i \omega_{n}t} \;,\\
    Q(z,t) &= \sum_{n} Q_n(z) e^{i \omega_{n}t} \;,
\end{align}
where \(\omega_{n} = \dfrac{2 \pi}{T} n\), with \(T\) being the cardiac cycle length. The number of terms was set to \(n=10\). Eq.~\ref{eq:PCNDE_cont} can be rewritten using the Fourier-series fit as:
\begin{align}
    \sum_{n}i\omega_{n}S_{n}(z)e^{i\omega_{n}t} = \sum_{n} - a(z) \dfrac{\partial Q_{n}(z)}{\partial z} e ^{i\omega_{n}t} \;.
\end{align}

Knowing the flow rate values from Eq.~\ref{eq:PCNDE_mom}, the Fourier coefficients for the area can be expressed as
\begin{align}
    S_{n} = - \dfrac{a(z)}{i \omega_{n}}\dfrac{\partial Q_{n}}{\partial z} \hspace{10pt} \mathrm{for} \hspace{10pt} n\neq0 \;.
\end{align}

The final step is to calculate the coefficient \(S_{0}\), using the initial condition from the first prediction:
\begin{align}
    S_{0} = \hat{S}(z,t=0) - \sum_{n\neq0}S_{n} + b(z) \;,
\end{align}
where \(\hat{S}(z,t)\) is the intermediate prediction from Eq.~\ref{eq:PCNDE_cont}.
In the Fourier series formulation, \(a(z)\) and \(b(z)\) are a new set of optimizable parameters, similar to \(\hat{a}(z)\) and \(\hat{b}(z)\) in the previous stage. These parameters are again fit using an MSE loss for \(S\). The resulting area \(S\) is fed back to the neural network \(f_{\theta}(Q,S)\) and the process is iterated. The algorithm is summarized in Fig.~\ref{fig:methods}.

\subsubsection{Pressure}
Due to inherent limitations of the 1D equations, the traditional 1D FEM models cannot capture the pressure drop along a stenosis (blocked artery). To overcome this, most studies, e.g.,~\citep{pfaller2022automated, grande20221d, blanco2018comparison}, use an empirical formula for the pressure drop along the stenosis. The most commonly used formula is based on the work of~\citet{seeley1976effect} where the pressure drop is written as:
\begin{align}
    \label{eq:Deltap}
    \Delta p &= \rho \dfrac{Q^{2}}{S_{D}^2} \dfrac{K_{v}}{\mathrm{Re}} + K_{t} \rho \dfrac{Q^{2}}{2 S_{D}^{2}} \left(\dfrac{S_{D}}{S_{s}} -1  \right)^{2} \;,\\
    K_{v} &= 32 \dfrac{L_{s}}{D_{0}}\dfrac{S_{D}^{2}}{S_{s}^{2}} \;, 
\end{align}
where \(S_{s}\) is the minimal cross-sectional area of the stenosed segment, \(L_{s}\) is the length of the stenosis, \(S_{D}\) is the healthy cross-sectional area upstream of the stenosis, \(\mathrm{Re}\) is the Reynolds number, \(D_{0}\) is the diameter of the healthy cross-section, and \(K_{v}\) and \(K_{t}\) are empirically defined parameters~\citep{seeley1976effect} ($K_{t}= 1.52$). This formulation will be used as the base of the pressure model here.

A two-stage pressure model was fit to predict the spatio-temporal variations of pressure. First, a small neural network was used to capture the temporal variations of the normalized pressure \(p^{*}=p/p_{ref}\):
\begin{align}
    \label{eq:p_NN}
    \hat{p}^{*}(z,t) = 1 + \dfrac{1}{p_{ref}}\dfrac{Eh}{r_{0}} g_{\theta}\left(S,S-S_{u},\Delta p, Q\right),
\end{align}
where \(\Delta p\) is calculated using Eq.~\ref{eq:Deltap}, \(r_{0}=D_{0}/2\), \(S\) and \(Q\) are the outputs of the previous neural PDE model, and \(g_{\theta}\) is a neural network with 4 layers of size \(4n_{z},2,2,n_{z}\) and \edit{tanh} activation functions. The form of Eq.~\ref{eq:p_NN} is inspired by the 1D constitutive models. The inputs to NN are the predicted area \(S\), the change in area compared to the undeformed area \(S-S_{u}\), the estimated pressure drop \(\Delta p\) based on Eq.~\ref{eq:Deltap}, and the predicted flow rate \(Q\). Each input is of size \(n_{z}\), therefore the input layer of the NN is of size \(4n_{z}=404\). The NN output is an unscaled pressure approximation of size \(n_{z}=101\), the same as the spatial size of \(p^{*}\). The pressure model is fit for all time instances at once where time instances are organized in a batch dimension using a full batch. Instead of trying to fit the absolute pressure \(p\), the normalized pressure \(p^{*}=p/p_{ref}\) is used, which makes the NN training easier. The total number of parameters in \(g_{\theta}\) is 1119. The neural network is trained with an MSE loss between the predicted normalized pressure \(\hat{p}^{*}\) and the normalized ground-truth \({p}^{*}_{GT}=p_{GT}/p_{ref}\) from 3D averaged data.

A second-stage model is fit to capture the pressure drop along the stenosis and model the spatial variations:
\begin{align}
    \label{eq:p_2ndfit}
    p^{*}(z,t) = \begin{cases}
\overline{\hat{p}^{*}(z,t)} + a_{1} & \text{if $z<= \dfrac{1}{3}L$} \;, \\
p^{*}\left(z=\dfrac{1}{3}L,t\right) - \left(\dfrac{\widetilde{\Delta p}}{p_{ref}} + b_{1}(t)\right)\tanh{\left(10 b_{2}(t) \left(z-2\right)\right)} & \text{if $\dfrac{1}{3}L< z <= \dfrac{2}{3}L $} \;, \\
p^{*}\left(z=\dfrac{2}{3}L,t\right) & \text{if $z > \dfrac{2}{3}L$} \;,
\end{cases}
\end{align}
where \(\widetilde{\Delta p}\) is the same as \(\Delta p\), except \(K_{t} = 1.52 + a_{2}\), \(\hat{p}^{*}\) is the first-stage prediction from Eq.~\ref{eq:p_NN} and \(\overline{()}\) denotes spatial average. $L$ is the length of the model and the stenosis is located at  \(z=2\) cm.  In the above equations, \(a_{1}\) and \(a_{2}\) are trainable scalar parameters, while \(b_{1}(t)\) and \(b_{2}(t)\) are trainable vectors of size \(n_{t}\). This corresponds to a simple constant-tanh-constant model for the spatial distribution of the pressure. It overcomes any smoothness issues arising from then neural network fit due to the different spatio-temporal scales. The empirical formula by~\citet{seeley1976effect} gives a reasonable estimate for the pressure drop, therefore it is used as starting point with additional learnable correction terms. The model is centered around the stenosis, such that the stenosis is at the middle of the spatial domain. The pressure drop approximation can be fit locally, with the terms inside the \(\tanh\) function controlling the slope of the drop. In the current cases, the stenosis is always at \(z=2\) cm, therefore this information is hardcoded inside the \(\tanh\) function. This limitation could be addressed in the future with a larger training set where the stenosis location is also learned. The parameters in the second stage are also fit with an MSE loss.

\subsection{Training Data}
\label{sec:data}
\begin{figure}[h]
    \centering
    \includegraphics[width=0.7\linewidth]{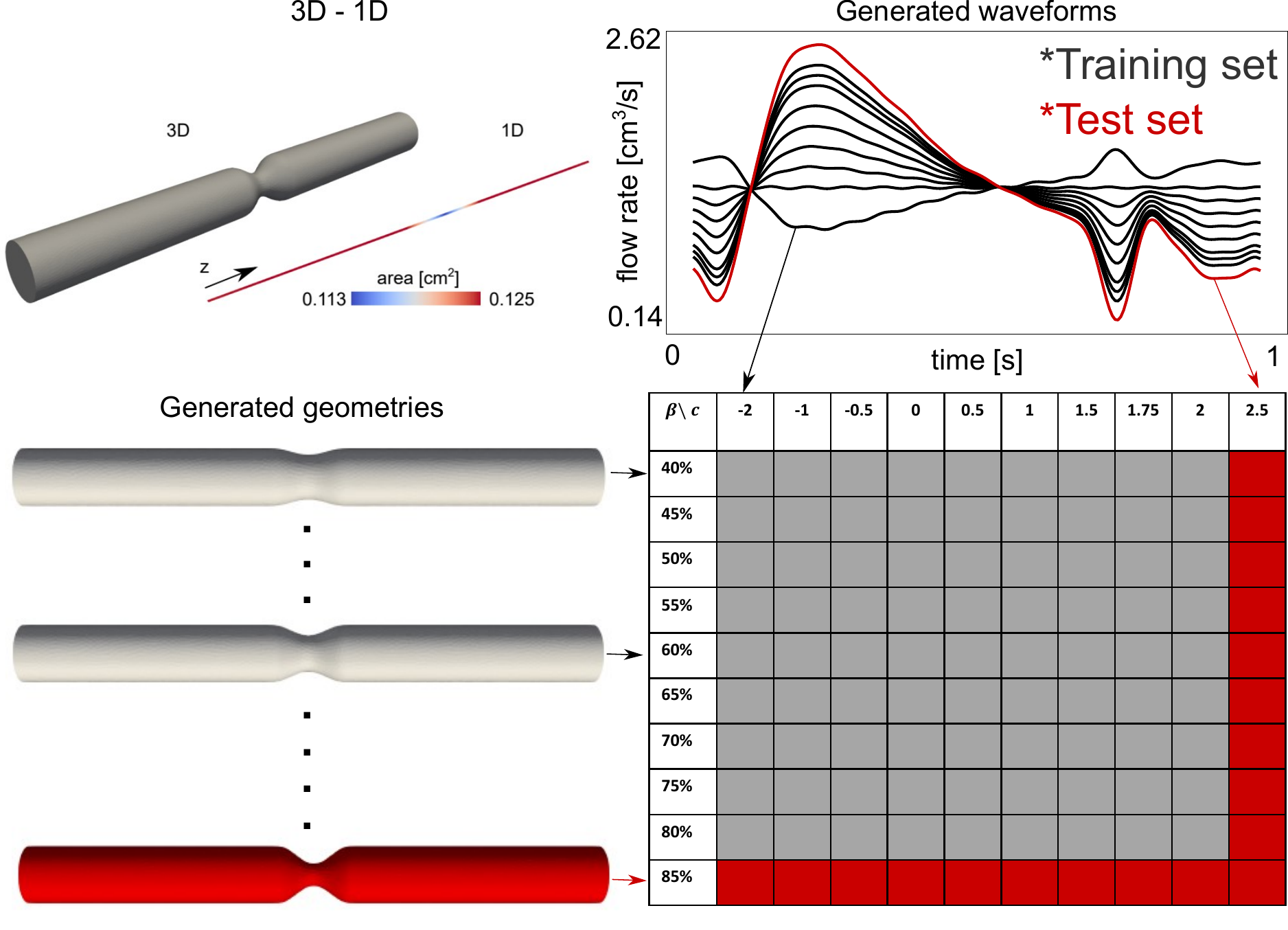}
    \caption{An overview of the training and test (extrapolation) datasets. The generated inlet waveforms are shown at the top, while the generated stenoses geometries are shown on the left. The 10 by 10 matrix represents the dataset, where gray cells are training data and red cells are test (extrapolation) data points.}
    \label{fig:train-test}
\end{figure}

To train the PCNDE model, 100 different 3D FSI simulations were run using SimVascular. There were 10 geometries with different stenosis blockage ratio \(\beta\) and 10 different inlet flow rate boundary condition waveforms. The reference boundary condition was taken from~\citep{kim2010patient}. Further waveforms were generated using a Fourier series fit
\begin{align}
    Q_{in}(t) = A_{0} + \sum_{1}^{n}\left ( c A_{n} \cos \left(\dfrac{2\pi n t}{T} \right) + B_{n} \sin \left(\dfrac{2\pi n t}{T} \right) \right) \;,
\end{align}
where \(c \in \left\{-2,-1,-0.5,1,1.5,1.75,2,2.5  \right\}\) is a fitting coefficient for generating more waveforms, while \(A_{n}\) and \(B_{n}\) are the Fourier coefficients. The number of terms was set to \(n=60\). The stenosis blockage ratio \(\beta\) was varied between 40\% and 85\%, with increments of 5\%. This led to the generation of a total of 100 CFD simulations with 10 different inlet boundary conditions and 10 different stenosis blockage ratios. These are shown in Fig.~\ref{fig:train-test}, in a data matrix, where each cell is one simulation case. The grey cells correspond to the training set, while the red cells to the test set. Each row of the data matrix corresponds to stenosis blockage ratios, while columns to inlet boundary condition waveforms. As seen, the test set is chosen to be outside the training regime, corresponding to extrapolation. The test geometry has the highest blockage ratio and the highest peak \(\mathrm{Re}\) number for the inlet waveform. The training set consisted of 81 cases, while the test set had 19 cases.

The geometry size is on the order of a left anterior descending (LAD) artery, a location notably prone to stenosis formation. The inlet diameter of the vessel was set to \(D_{in} = 4\) mm, while the total vessel length was \(L= 40\) mm, with the domain being centered around the stenosis. The idealized stenosis shape was generated according to the analytical formula from~\citet{sherwin2005three}.

The coupled momentum method~\citep{figueroa2006coupled} was used for modeling the elastic wall behavior in 3D, while a linear pressure-area constitutive relationship was used in 1D. The wall thickness was set to \(h_{w}=0.5\) mm, Young's modulus of the wall material to \(E = 1.5\) MPa, the blood density to \(\rho = 1060 \) kg/m\(^{3}\), and the dynamic viscosity to \(\mu = 0.004\) kg/ms, for both 1D and 3D models. For the 3D model, the Poisson's ratio of the vessel wall was set to \(\nu_{w} = 0.5\), the wall density to \(\rho_{w}=1075\) kg/m\(^{3}\), and the shear constant to \(k_{w}=0.83\). The outlet boundary condition was a resistance-type Windkessel element with a reference pressure of \(p_{ref} = 100\) mmHg for both 1D and 3D in all cases. The 3D cases all had an unstructured mesh of around 426k--433k elements, using a boundary layer mesh near the vessel wall. The cardiac cycle length was 1 s and the simulation timestep was set to be \(\Delta t = 10^{-3}\) s. The cases were run for 3 cardiac cycles, and results were saved from the last one at \(n_{t}=100\) equispaced time instances. After the 3D simulations were done, results were cross-sectionally averaged at \(n_{z}=101\) spatial locations, at increments of \(0.04\) cm from the inlet to the outlet. The 3D FSI simulations took, on average, 3-4 hours on 24 processors at the University of Utah CHPC cluster.

Tsit5~\citep{tsitouras2011runge} was used as the ODE solver from the \textit{DifferentialEquations.jl}~\citep{rackauckas2017differentialequations} Julia library to solve the neural PDE problem with an adaptive timestep. \edit{Tsit5 is a 5th order Runge-Kutta method, with an embedded 4th order method for estimating the error. The error estimate is used for the adaptive time stepping~\citep{rackauckas2017differentialequations}.} Mean squared error (MSE) loss functions were used between the predictions and the ground-truth 3D averaged data. All models were trained with the BFGS optimizer. The batch size was set to 12 for the flow rate and area models, and to a full batch for the pressure model. The neural network in the momentum equation was trained for 100 epochs with 5 iterations inside each batch, for a total of 500 iterations. The area coupling was done every 25 epochs. This approach enhances stability by allowing the momentum equation's neural network to train for several epochs before coupling. Additionally, since the continuity equation converges more rapidly, more frequent coupling is unnecessary. The parameters in the continuity equation were optimized for 5 iterations for each batch. The neural network \(g_{\theta}\) in the pressure equations was trained for 250 iterations, while the second stage scalar (\(a_{1},a_{2}\)) and vector (\(b_{1},b_{2}\)) parameters for 75 iterations. In inference mode, the model is run for 3 iterations, the area predictions are coupled back to the momentum equation after each one. The total cumulative number of optimizable parameters in the PCNDE model was 4449, which was trained and run on a single Intel Xeon W-1370P CPU.

SimVascular's svOneDSolver was used to solve the 1D blood flow equations using 1D finite elements. This was only done for the 19 cases in the test set. This serves as a benchmark for our PCNDE predictions. Inlet and outlet boundary conditions and material parameters were set up to be the same as the 3D model. The 1D mesh had 400 elements along the centerline, with a timestep size of \(\Delta t = 10^{-4}\)s, and the simulation was run for three cardiac cycles. The results were saved at \(n_{t} = 100\) time instances from the third cycle. The convergence tolerance was set to \(10^{-8}\). 

It is important to note that since the data comes from 3D averaged simulations, it does not exactly satisfy the 1D governing equations (Eq.~\ref{eq:cont} and~\ref{eq:momentum}). This suggests that adding further physics terms on the right-hand side of Eq.~\ref{eq:PCNDE_mom} is not strictly beneficial. This is further investigated in Section~\ref{sec:loss_landscapes}.

\section{Results}
\label{sec:results}
\subsection{Temporal Model}
\label{sec:temporal_NPDE_results}
\begin{figure}[h]
    \centering
    \includegraphics[width=\linewidth]{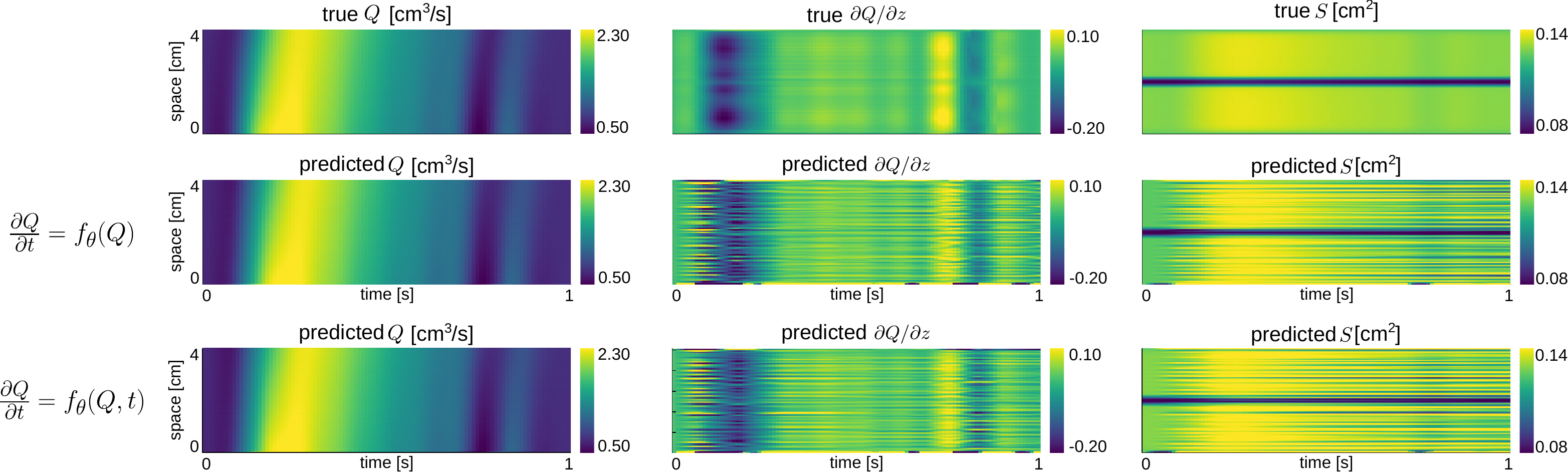}
    \caption{Temporal neural PDE results. The left panel shows the true and predicted flow rate. The middle panel shows the true and predicted spatial derivatives of the flow rate. The right panel shows the true and predicted area values. \edit{The top row shows the ground-truth 3D averaged data, the middle rows shows the model results where only the flow rate is input to the NN (\(f_{\theta}(Q)\)), the bottom row shows the model results where both flow rate and time are inputs to the NN (\(f_{\theta}(Q,t)\)).}}
    \label{fig:dQdt}
\end{figure}

Example results from the temporal formulation are shown in Fig.~\ref{fig:dQdt}. \edit{Two models are investigated, one where the NN only takes the flow rate as input, i.e., \(f_{\theta}(Q)\). Results from this model are shown in the middle row. Results from a second model are shown in the last row, where the NN takes as input time too, i.e., \(f_{\theta}(Q,t)\).} The left panel shows the true and predicted flow rates as a function of space and time. It can be seen that the flow rate predictions are quite accurate. Nevertheless, the spatial derivative of the flow rate \(\partial Q/\partial z\) shows a non-smooth behavior, seen in the middle panel. This non-smoothness is then transferred to the area predictions, seen in the right panel. Therefore, the temporal formulation poses challenges with coupling the results of the neural PDE output for flow rate, with the continuity equation. The non-smoothness issue is carried over to the area predictions because the spatial derivatives of the flow rate directly show up in the continuity equation. In more extreme cases, this non-smoothness can also cause the \edit{ODE solver} to diverge. \edit{It can also be observed that introducing time as an additional input to the NN does not change the results significantly.}
\subsection{PCNDE results}

\begin{figure}[htbp]
    \centering
    \includegraphics[width=0.95\linewidth]{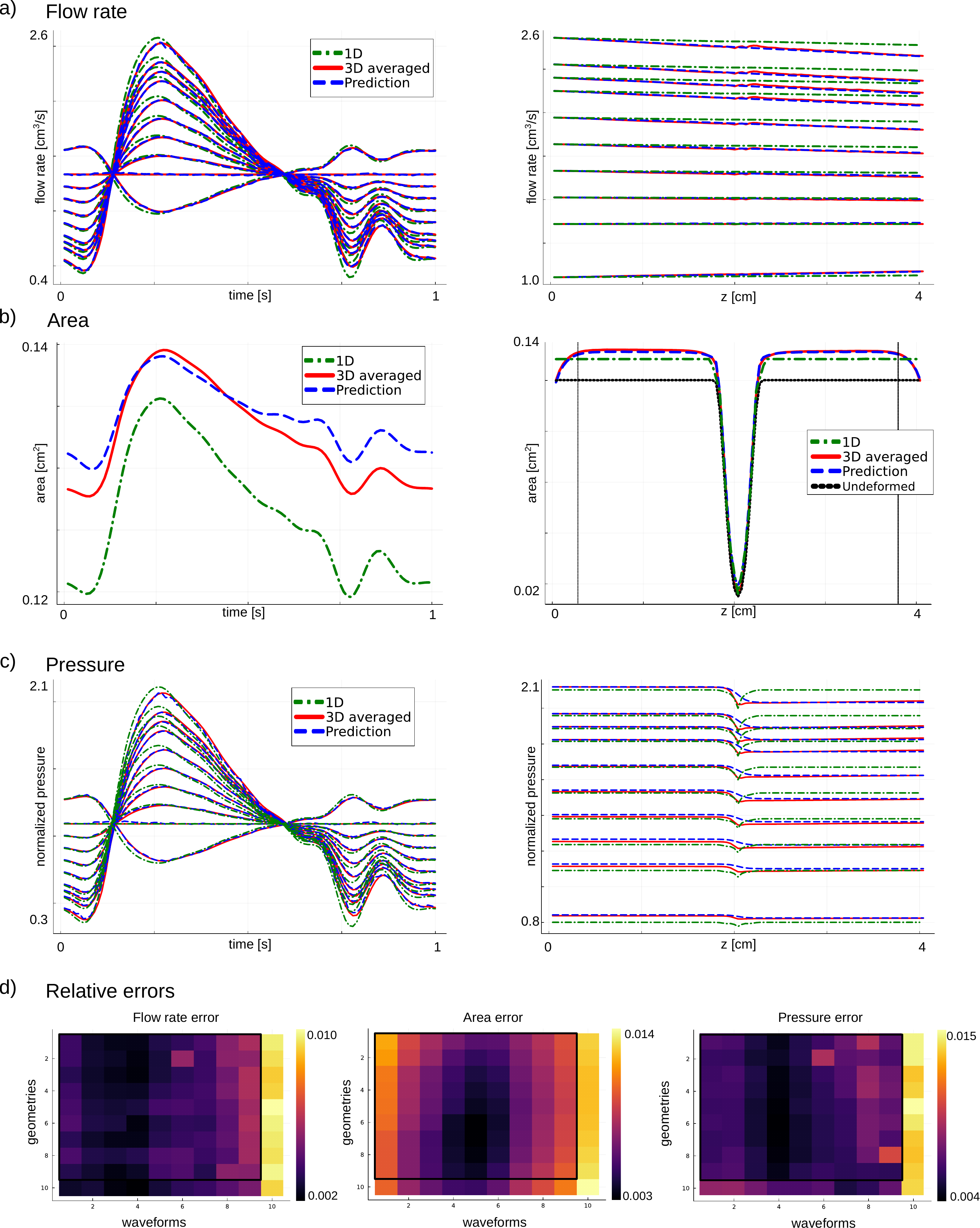}
    \caption{PCNDE results compared with the 1D FEM results. Panel a) shows the flow rate \(Q\) results as a function of time and space. Panel b) shows the cross-sectional area \(S\) results as a function of time and space. For better visibility, only one case is shown for the area. \edit{The errors are only calculated between the two black vertical lines to exclude boundary effects.} Panel c) shows the normalized pressure \(p/p_{ref}\) results as a function of time and space. The plots as a function of time are shown at \(z=3.6\) cm, while the plots as a function of space are shown at \(t=0.25\) s. Panel d) shows the PCNDE relative errors for all three variables of interest for the 100 different cases.}
    \label{fig:results}
\end{figure}

\begin{figure}[htbp]
    \centering
    \includegraphics[width=\linewidth]{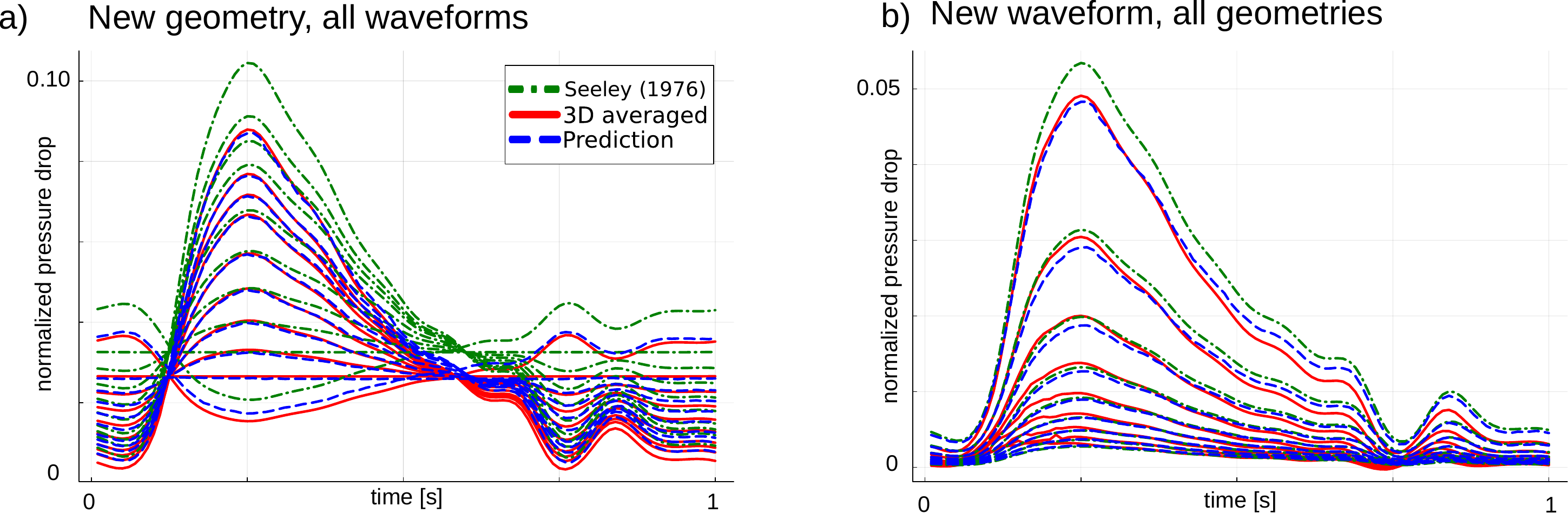}
    \caption{Pressure drop predictions are shown. a) Pressure drop for the new geometry in the test dataset with all waveforms (bottom row of the data table in Fig.~\ref{fig:train-test}). b) Pressure drop for the new waveform in the test dataset with all geometries (right column of the data table in Fig.~\ref{fig:train-test}).}
    \label{fig:deltaP}
\end{figure}

The flow rate, cross-sectional area, and pressure results for the PCNDE model are shown in Fig.~\ref{fig:results}. Figure~\ref{fig:results}a shows the flow rate as a function of time and space for 10 waveforms from the test dataset with the new geometry (bottom row of the data table in Fig.~\ref{fig:train-test}). The PCNDE predictions (blue) are compared to the 1D FEM solution (green) and the 3D averaged ground truth data (red). For all panels, the temporal plots \edit{(left column)} are shown at \(z=0.9L=3.6\) cm, an arbitrary location downstream of the stenosis, close to the end of the domain, while the spatial plots \edit{(right column)} are shown at \(t=0.25\) s, corresponding to the peak flow rate. The 1D FEM captures the overall temporal evolution well, but it deviates from the 3D averaged solution around peak flow rate. However, the PCNDE prediction captures the dynamics better in both time and space. It is noteworthy that since the simulations include moving walls, the flow rate along the vessel is not constant. Figure~\ref{fig:results}b shows the area variations as a function of time and space for the new geometry and new waveform (bottom right cell of the data table in Fig.~\ref{fig:train-test}). The 1D FEM underestimates the area deformations, and the PCNDE model captures the area variations well around the largest deformations but overestimates them at the beginning and the end of the cardiac cycle. Note that the PCNDE predictions are explicitly periodic due to the Fourier series fit. This will be further discussed in Sec.~\ref{sec:fourier}. Figure~\ref{fig:results}c shows the pressure as a function of space and time for the same part of the test dataset as for the flow rate. Overall, the trends are similar, and the PCNDE model captures the pressure variations in time better than the 1D FEM, especially around the highest pressures. \edit{There are minor oscillations in the temporal pressure predictions, particularly around the peak of the unseen waveform. While these oscillations are relatively small, they warrant caution when extrapolating beyond the training set.} It can also be seen that the simple spatial model captures the pressure variations well, even though the slope of the pressure drop is usually underestimated. It is clear that the 1D FEM model cannot capture the pressure drop well. The normalized pressure drop \(\Delta p/p_{ref}\) obtained from the PCNDE model is compared to the empirical formula of~\citet{seeley1976effect} in Fig.~\ref{fig:deltaP}. Panel a) shows the comparison for the new geometry (bottom row of the data table in Fig.~\ref{fig:train-test}), while panel b) shows the new waveform (right column of the data table in Fig.~\ref{fig:train-test}) as a function of time. For both cases, the PCNDE prediction is closer to the 3D averaged ground truth. The empirical formula captures the shape of the pressure drop evolution but overestimates the pressure drop in most cases. \edit{The neural network in the pressure model, \(g_{\theta}\), has been tested with multiple activation functions, including GeLU, ReLU, tanh, sigmoid, and swish. The network was robust with respect to the choice of activation function, with GeLU, tanh, sigmoid, and swish producing results that differed in their relative train and test errors less than 0.001. The only exception was ReLU, which failed and only learned the temporal mean.}

\subsection{Error Analysis}

\begin{figure}[h]
    \centering
    \includegraphics[width=\linewidth]{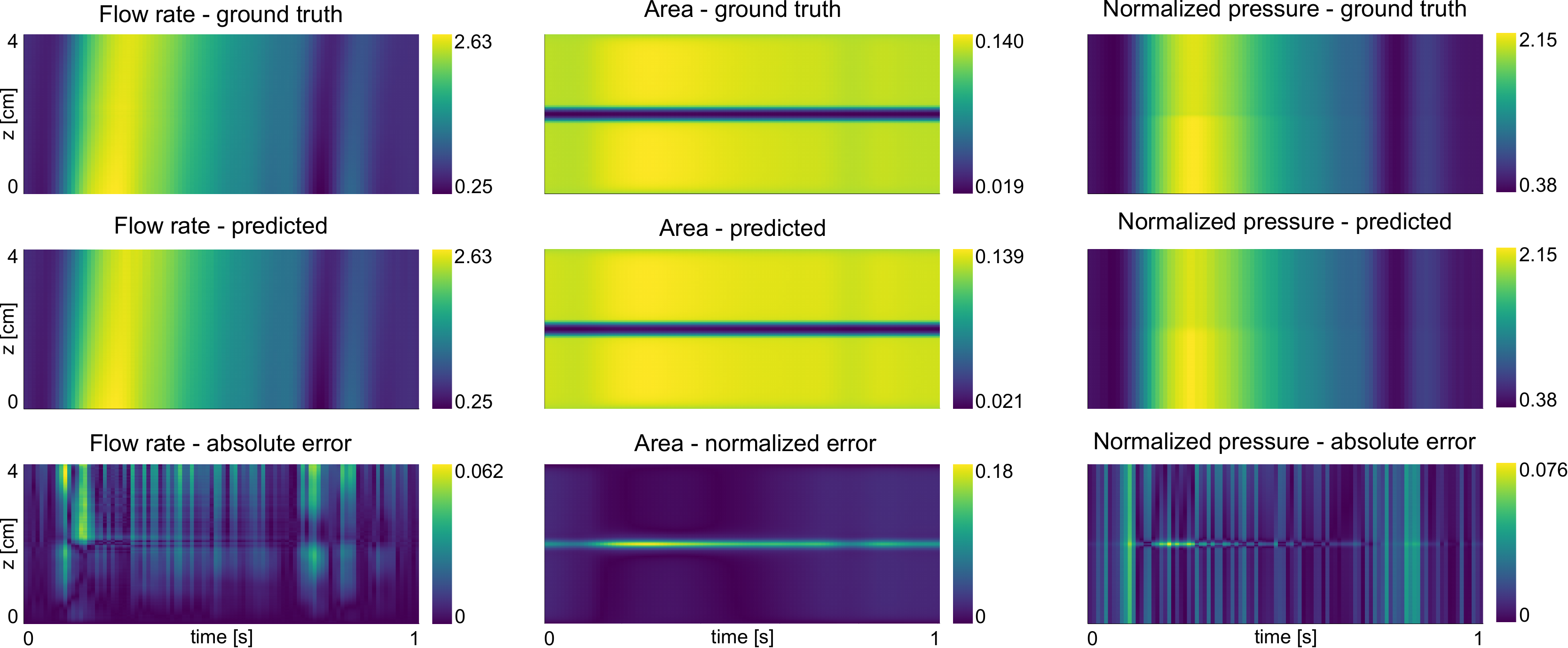}
    \caption{Flow rate, area, and normalized pressure comparisons for the new waveform and new blockage ratio from the test dataset. The top row shows the 3D averaged ground-truth values, the middle row shows the PCNDE predictions, and the last row shows the absolute point-wise errors. \edit{For area, the absolute errors are normalized by the area at \(t=0\) to highlight the differences, since the area variations are relatively small.}}
    \label{fig:pointwise_error}
\end{figure}

Results are plotted as a function of space and time in Fig.~\ref{fig:pointwise_error} for the extrapolation case with the new waveform and new stenosis blockage ratio. The top row shows the 3D averaged ground-truth values, the middle row shows the PCNDE predictions, and the bottom row shows the point-wise absolute errors. \edit{For area, the absolute errors are divided by \(S(t=0)\) for normalization as the area changes are relatively small.} The overall patterns match well for all three variables of interest. The absolute error plot for flow rate shows a staggered pattern, but on average, the errors are higher in the second part of the domain, behind the stenosis. For area, the highest errors clearly arise around the stenosis region in space and in the first half of the cardiac cycle. \edit{The normalized errors outside the stenosis are very small.} For pressure, a staggered error pattern emerges in time, with mostly uniform distribution in space.

The relative errors for the training and test sets are reported in Table~\ref{tab:result} for all three variables, along with the 1D FEM relative errors for the test set. To account for the boundary effects in the 1D and 3D FEM models, the area errors do not take into account the first and last 5 elements. The PCNDE model has around 3-5 times smaller errors than the 1D FEM solution. We emphasize that all the test cases are outside the training parameter regime, therefore, considered extrapolations. For both the PCNDE and the 1D FEM models, the flow rate has the lowest errors among the three variables of interest.  For a more detailed analysis of the error distribution among different cases, Fig.~\ref{fig:results}d shows the PCNDE relative error for all three variables of interest of all 100 cases. The training set is distinguished with a black square and cases outside the black square are the test set. For all variables, it is evident that the new waveform shape has a higher error than the new stenosis blockage ratio, i.e., the last column has the highest errors. The bottom row has similar errors to the training set for all cases. For area, the smallest errors are in the middle of the training parameter regime. For flow rate and pressure, the smallest errors are around waveforms 2-4, corresponding to waveforms with the lowest temporal variability. For area, the second highest error is for the first column, corresponding to the first waveform. This waveform has a minimum where all others have their peak, therefore it is a more challenging waveform for the algorithm. Overall, the worst-case error for the flow rate is around 1\%, 1.4\% for area, and 1.7\% for pressure.

\begin{figure}[h!]
    \centering
    \includegraphics[width=\linewidth]{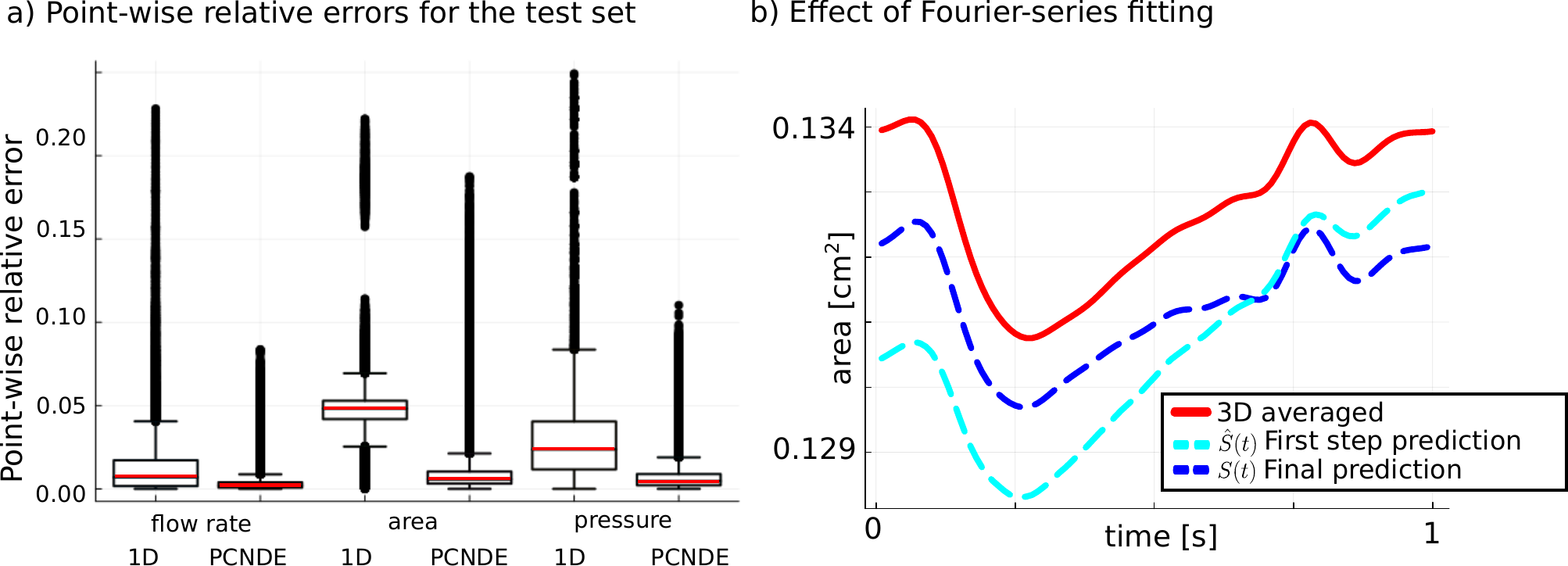}
    \caption{a) Box plots for the relative point-wise errors for flow rate, area, and pressure. Both 1D and PCNDE results are shown. The red line represents the median, the edges of the box indicate the lower and upper quartiles, whiskers extend to the non-outlier minimum and maximum values, and any outliers are represented as individual points beyond the whiskers. b) The effect of the Fourier series fitting for the area for the first waveform from the training dataset is shown as an example.}
    \label{fig:box_plots}
\end{figure}

Box plots of point-wise relative errors are shown in Fig.~\ref{fig:box_plots}a for all points from the test set for both the 1D FEM and PCNDE results. The red lines represent the median. In addition to having a lower median error, the PCNDE results also have a smaller spread with smaller outliers. For flow rate and pressure, all PCNDE errors are less than 10\%, while for the area the largest outlier is around 19\%. On the other hand, in the 1D FEM, the largest outliers are above 20\% for all variables. For 1D FEM, area has the highest median error, however, the pressure has the largest outliers. For PCNDE, pressure and area have similar median error, but the area has the largest outliers. 

\begin{table}[h]
  \centering
  \begin{tabular}{c|c|c|c}
    \multicolumn{4}{c}{Relative errors: \( \left( \Vert Y -  Y_{GT}  \Vert \right)  / \ \Vert Y_{GT} \Vert\)}                   \\
    \hline
    Variable     & Training - PCNDE     & Test - PCNDE & Test - 1D FEM \\
    \hline
    Flow rate \(Q\) & 0.0040  & 0.0075  & 0.0204   \\
    Area \(S\)    & 0.0077 & 0.0114 & 0.0498  \\
    Pressure \(p\) &\edit{0.0080}  & \edit{0.0109} & 0.0365\\
  \end{tabular}
  \caption{PCNDE and 1D FEM relative errors for flow rate, area, and pressure. For 1D FEM, only the cases in the test set are reported.}
   \label{tab:result}
\end{table}

\subsection{Enforcing Temporal Periodicity with Fourier Series}
\label{sec:fourier}

To observe the effect of the additional Fourier series fit for the cross-sectional area predictions, Fig.~\ref{fig:box_plots}b shows an example of the first prediction \(\hat{S}(t)\) from Eq.~\ref{eq:PCNDE_cont} (cyan) and the final prediction \(S(t)\) after the Fourier series fit (blue) for the first waveform from the training set. The first step prediction provides a reasonable estimate for the area, however, it clearly violates temporal periodicity. The subsequent Fourier series fit, leading to the final prediction, is periodic by design. In terms of error, in some regions, the final prediction produces higher error than the first step prediction, for example, around the end of the cardiac cycle in this case. However, in most of the domain, the periodic prediction is closer to the ground truth, thus has a smaller error. The two-stage fitting process serves a dual purpose. Initially, it allows the prediction to transition smoothly from the undeformed area to a close approximation of the ground truth, albeit without adhering to temporal periodicity. Subsequently, the second stage refines this approximation by applying a Fourier series fit, effectively enforcing periodicity in the final prediction.

\subsection{Sensitivity to Noise in Inputs}
\label{sec:sensitivity_Qin}
\begin{figure}[h]
    \centering
    \includegraphics[width=\linewidth]{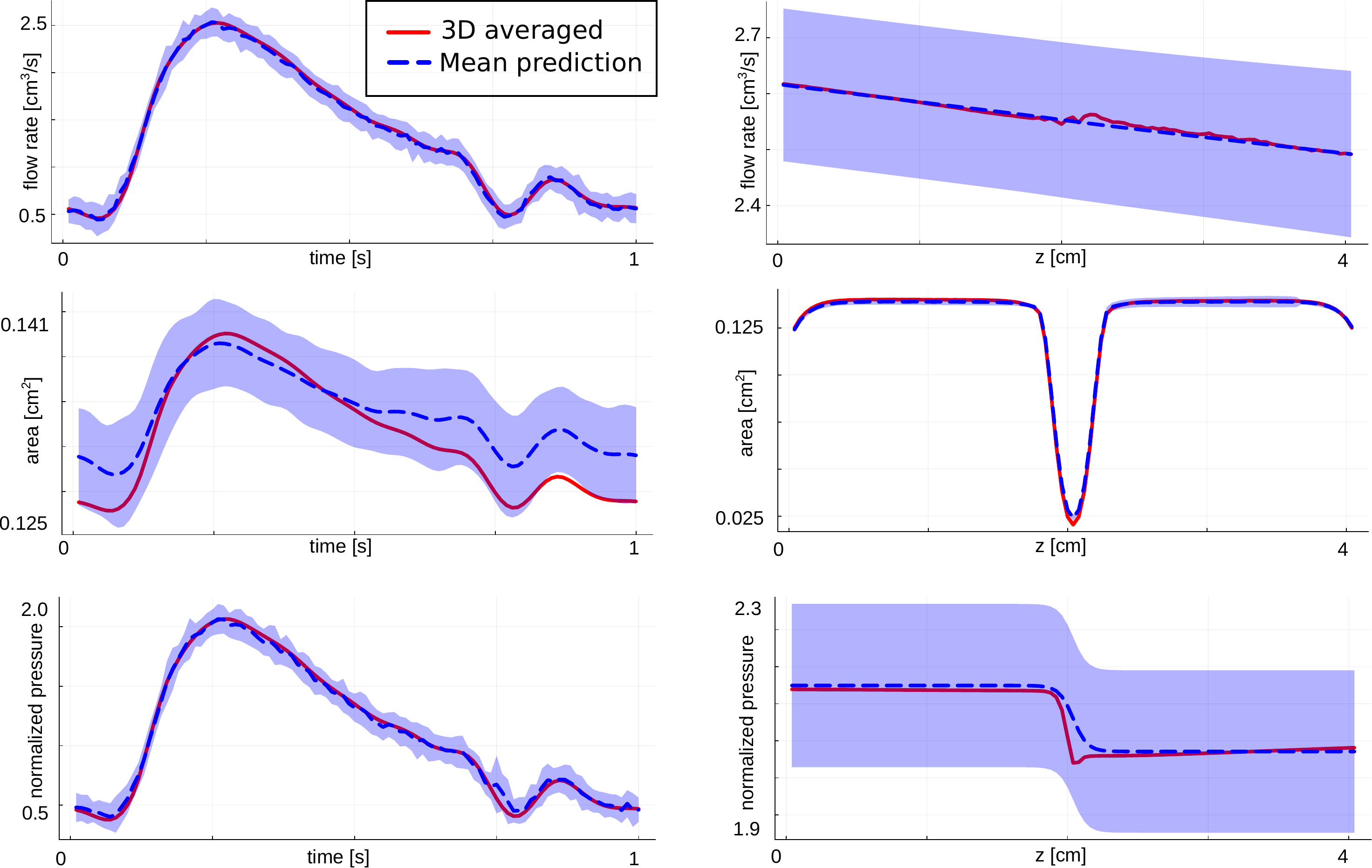}
    \caption{Uncertainty visualization given a noisy inlet boundary condition \(Q_{in}(t)\) is shown for flow rate, area, and normalized pressure results. The left column shows the temporal plots at \(z=3.6\) cm and the right column shows the spatial plots at \(t=0.25\) s. The dashed blue lines represent the mean of the predictions and the shaded regions correspond to the standard deviation of the predictions. }
    \label{fig:uncertainty_Qin}
\end{figure}

The two main inputs to the PCNDE model are the inlet boundary condition flow rate waveform \(Q_{in}(t)\) and the undeformed cross-sectional area \(S_{u}(z)\). To investigate the PCNDE model's sensitivity, noise was added to the inlet flow rate waveform:
\begin{align}
    Q_{in}^{*}(t) = Q_{in}(t) + \varepsilon \sigma \;,
\end{align}
where \(\varepsilon \sim \mathcal{N}(0,1) \) is the standard normal distribution, \(\sigma = 0.05 \max(Q_{in}(t))\) is the noise level scaled by the maximum of the waveform. 90 samples were taken from the random distribution and the model was run in inference mode for the extrapolation test case of a new waveform and new geometry. The results are shown in Fig.~\ref{fig:uncertainty_Qin} for flow rate, area, and normalized pressure as a function of time and space. The temporal plots are at \(z=3.6\) cm, and the spatial plots at \(t=0.25\) s. The dashed blue lines represent the means of the predictions, while the shaded regions correspond to the standard deviation of the predictions. The mean standard deviation relative to the maximum flow rate was \(0.052\), which is similar to the amount of noise added to the inlet boundary condition waveform. The mean prediction fits well the 3D averaged ground-truth data, both in space and time. Note that the ranges for the vertical axes are different for the temporal and spatial plots. The mean standard deviation of the area predictions relative to the maximum cross-sectional area was \(0.014\), indicating that the errors are not exaggerated during the area prediction. A higher uncertainty is observed in the region after the stenosis for area. Smaller uncertainty is observed in time before the peak flow rate, around 0.1-0.25 s for all variables. Apart from that, the level of uncertainty remains relatively constant over time for all variables. The mean standard deviation of the normalized pressure relative to the maximum was \edit{\(0.058\)}, which is similar to what was observed for the flow rate. A further sensitivity study is included in the Appendix, Sec.~\ref{sec:sensitivity_Su}, regarding the undeformed cross-sectional area.


\subsection{Including Additional Physics Terms and Visualizing Loss Landscapes}
\label{sec:loss_landscapes}
\begin{figure}[h]
    \centering
    \includegraphics[width=\linewidth]{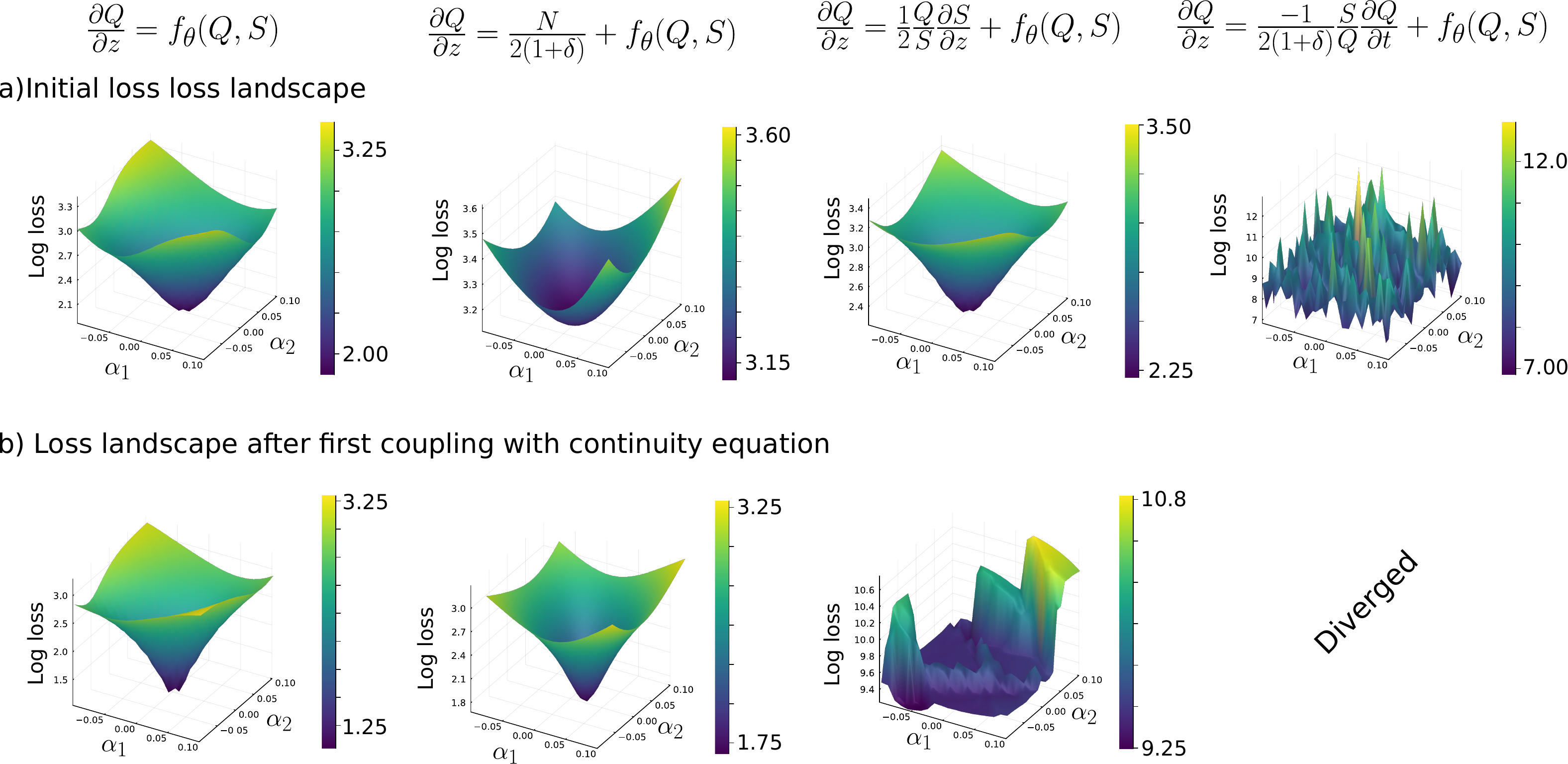}
    \caption{Loss landscapes of the momentum equation for different physics terms included. The model equations are shown at the top of each column. a) The top row shows the initial loss landscape during the first phase of training. b) The bottom row shows the loss landscape after the first coupling with the continuity equation. }
    \label{fig:loss_landscape}
\end{figure}

In this section, we explore including more physics terms on the right-hand side of Eq.~\ref{eq:PCNDE_mom}. It is observed that the additional physics terms can make the problem more challenging from the optimization point of view. After expanding and reordering Eq.~\ref{eq:momentum} to have \(\partial Q/\partial z\) on the left-hand side, the following form can be achieved:
\begin{align}
    \dfrac{\partial Q}{\partial z} = \dfrac{1}{2} \dfrac{Q}{S} \dfrac{\partial S}{\partial z} + \dfrac{N}{2(1+\delta)} - \dfrac{1}{2(1+\delta)} \dfrac{S}{Q} \dfrac{\partial Q}{\partial t} + f_{\theta}(Q,S).
    \label{eq:mom_physics}
\end{align}
To analyze the effect of the inclusion of different physics-based terms on the right-hand side of Eq.~\ref{eq:mom_physics}, the loss landscapes were investigated for four cases. The spatial and temporal derivatives on the right-hand side were taken using a second-order central difference scheme. The visualization of loss landscapes was based on~\citep{subel2023explaining,li2018visualizing}. Assume that \(\theta \in \mathbb{R}^{p}\) are the trainable parameters in \(f_{\theta}\), with \(p=2724\) in this case. To visualize a 2D surface of the loss function, two random direction vectors \(v_{1}, v_{2} \in \mathbb{R}^{p}\) in parameter space are selected. Then, the network parameters are perturbed by adding \(v_{1}\) and \(v_{2}\) 
\begin{align}
    \theta_{p} = \theta + \alpha_{1} v_{1} + \alpha_{2} v_{2} \;,
\end{align}
where \(\theta_{p}\) are the perturbed parameters, and \(\alpha_{1}, \alpha_{2}\) are the amplitudes for the random perturbations. The loss can be computed for all amplitudes in the predefined range of \(\alpha_{1}, \alpha_{2} \in [-0.1,0.1]\). The loss landscape is plotted as a surface, being a function of \(\alpha_{1}\) and \(\alpha_{2}\). The loss landscapes for four different models are shown in Fig.~\ref{fig:loss_landscape} with different terms on the right-hand side. The top row shows the loss landscape after five BFGS iterations, while the bottom row shows the loss landscape after the first coupling of the predicted \(S\) to \(f_{\theta}(Q,S)\) after solving the continuity equation. The different equations are shown at the top of each column. The first column shows the case where the right-hand side only has a neural network. Both loss landscapes are smooth and have a clear minimum. The second column, where a constant term is added, shows similar loss landscapes; however, the loss values are higher than those in the pure NN case. The third column presents a case where the \(\partial S /\partial z\) term is added. In this scenario, the initial loss landscape looks similar to the pure NN case. However, the obtained area solution is non-smooth and not accurate, which causes the loss landscape to change after the area is coupled back to the momentum equation. This case eventually diverges. The last column is with the \(\partial Q/ \partial t\) term, which has an extremely difficult loss landscape with multiple local minima, and the loss is high as well. This case diverges even before getting to the first iteration of the continuity equation solver.

These results justify the choice of using a pure neural network on the right-hand side of the momentum equation. Adding further physics terms can make the optimization problem much harder. Unlike classical machine learning, in this case, differential equations need to be solved, which have to be stable throughout all the training. Therefore, a certain level of model robustness is needed to make sure the integration will not diverge. This is a well-known problem with training neural networks inside physics solvers~\citep{sanderse2024scientific,um2020solver}. Also, note that the equations are not exact since the ground-truth data comes from 3D averaged simulations and not the 1D blood flow equations. Therefore, these terms likely do not represent the correct physics in their current form. Instead, using a standalone neural network allows us to exploit its flexibility and expressivity to learn the underlying physics, especially when there is uncertainty about the structure of the true physical terms.

\section{Discussion}
\label{sec:discussion}
The computational efficiency of 1D blood flow models makes them particularly attractive for clinical applications where timely results are essential.  However, there is still a discrepancy between 1D model predictions and cross-sectionally averaged results from more detailed 3D FSI simulations, particularly in disturbed blood flow environments. The proposed PCNDE framework addresses this gap by maintaining the computational speed of the 1D approaches, while reducing errors relative to the 3D averaged solutions. This novel approach strikes a balance between data-driven methods and physics-informed modeling. It incorporates prior knowledge through the continuity equation and a pressure model, while introducing learned correction terms to capture details that the 1D models may overlook. This approach is designed to enhance the accuracy of 1D models without sacrificing their computational efficiency, potentially bridging the divide between rapid clinical applicability and high-fidelity flow representation. \edit{The proposed PCNDE framework achieved less than 1.2\% relative error for both train and test cases for all variables of interest compared to the ground-truth 3D averaged FSI data. It also achieved 3-5 times smaller error than traditional 1D FEM for the test set, while having similar computational cost in inference mode.}

A significant challenge encountered during model development was the handling of coupling between the momentum and continuity equations. This requires careful handling to ensure stability and smoothness throughout all intermediate solution steps, even during the early stages of neural network training. The necessity for such stability and smoothness is essential, as it enables both the solution of the differential equation system and the backpropagation through the solver. These issues were particularly evident in our initial experiments with the temporal formulation, as detailed in Sec.~\ref{sec:temporal_NODE}, but also while trying to include further physics terms in Sec.~\ref{sec:loss_landscapes}. Analysis of the loss landscapes revealed significant variations in optimization dynamics across different terms, highlighting the critical role each component plays in shaping the model's performance. The quest for stability in differentiable simulators remains an active and challenging area of research~\citep{sanderse2024scientific}. To address these challenges, an innovative approach was adopted. Namely, reformulating the differential equations in the spatial domain rather than the temporal one, while leveraging the inherent temporal periodicity of blood flow problems. This strategic shift not only simplified the learning problem but also facilitated a smoother and more stable coupling between the equations.

While reformulating the equations in the spatial domain offered significant advantages, it should be acknowledged that this approach introduces its own set of challenges. The interchange of space and time variables fundamentally alters the nature of the differential equations, potentially affecting the representation of critical physical phenomena such as wave propagation. The physics terms in the momentum equation, when rewritten for the spatial formulation, become less intuitive and lose their clear physical meaning. However, this approach aligns well with the inherent temporal periodicity of blood flow problems, eliminating the need to solve for arbitrary future times, which is a stark contrast to most unsteady reduced-order modeling problems in fluid dynamics that aim for long-term temporal predictions beyond the training regime.

A key consequence of our spatial reformulation is the necessary reevaluation of boundary and initial conditions.  The space-time switching transforms the original inlet boundary condition into an initial condition, simplifying implementation within the Julia language framework. This transformation is beneficial because initial conditions are generally easier to handle than boundary conditions in this computational environment. The original flow rate initial condition was trivial and is omitted from the model, which is acceptable given the periodic nature of the temporal dynamics. While the current implementation does not explicitly address outlet boundary conditions, this aspect will become critical when extending the model to geometries with multiple outlets. It is important to note that while the momentum equation is solved in the spatial domain, the continuity equation remains in the temporal domain. This dual-domain approach introduces an interesting dynamic to the model. The continuity equation, being time-dependent, requires an initial condition, which we set as the undeformed area of the vessel. This split between spatial and temporal domains creates a unique coupling challenge where the momentum and continuity equations cannot be coupled after each step of their respective solvers, as one progresses through space while the other advances in time. Instead, the coupling can only occur after a complete solution of each equation. This approach to solving and coupling the equations represents a significant departure from traditional methods. It allows us to leverage the advantages of spatial formulation for the momentum equation while maintaining the temporal nature of the continuity equation. However, it also introduces complexities in ensuring consistency between the two domains. This method ensures that each equation can be solved efficiently in its most suitable domain, but it also means that the physical interactions between flow momentum and area changes are captured in a more discrete and iterative manner.

Another family of techniques that leverage the temporal periodicity of blood flow are frequency domain formulations of the 1D equations~\citep{papadakis2019wave, flores2016novel}. These formulations transform the governing equations from the space-time domain to the space-frequency domain using Fourier transform. By doing so, they enable both analytical and numerical solutions, particularly in idealized geometries such as straight vessels, simple bifurcations, or idealized stenoses. While these frequency domain models show promise, they have not yet been extensively validated on large, diverse datasets, leaving room for further investigation into their robustness and generalizability in real-world scenarios.

\edit{A fundamental modeling difference between the 3D and 1D models is that the inlet and outlet surfaces are fixed in 3D model. This leads to zero wall deformations in the 3D model, which is not the case in the 1D model. This is common practice and its effects are known to be negligible~\citep{xiao2014systematic}, except for the smaller displacements of the wall locally near the inlet and outlet. To mitigate this effect the area error calculations did not include the first and last 5 spatial points.}

An inherent challenge in cardiovascular flow modeling, which persists regardless of the chosen formulation, is the multiscale nature of the spatiotemporal variation in variables. Usually, the spatial and temporal scales of the problem differ by at least an order of magnitude. In the case of flow rate and pressure, temporal variations are much larger than spatial axial variations. On the other hand, for cross-sectional area, spatial variations are larger due to the presence of stenosis. Usually, fitting multi-scale phenomena with machine learning approaches is particularly challenging, and requires some additional model development~\citep{lutjens2022multiscale}. In the case of area and pressure, a two-stage fitting process is exploited to help with the multiscale phenomena, while for flow rate, the spatial variations are approximately linear and therefore easier to model. 
\edit{The second stage pressure model assumes that the location of the stenosis is known beforehand, based on the geometric information. Furthermore, we assume there is only one stenosis in the given artery. These limitations could be overcome by utilizing a larger and more diverse training data, containing arteries with various number of stenoses and different locations. The other promising option is to use a complementary tool for automated stenosis detection, e.g., similar to the one used by~\citet{pfaller2022automated}.}

The neural networks used in neural ODE/PDE approaches are typically small and relatively shallow~\citep{shankar2022validation,gelbrecht2021NPDE,lai2021structural, portwood2019turbulence, fedorov2023kinetics}, which contrasts sharply with traditional deep learning methods that often utilize much larger architectures. This design choice brings multiple benefits. First, it helps mitigate overfitting, which is crucial in these applications, given that the datasets are usually relatively small. Additionally, the reduced computational training costs are significant. Our model was trained on a single CPU in just a couple of hours. The small number of parameters also means that fewer iterations are needed to find the optimal solution during training, further enhancing efficiency. Lastly, in the context of ODEs and PDEs, smaller networks tend to produce more stable solutions, which is critical for accurately modeling coupled physical systems where numerical instabilities can easily lead to unrealistic results or divergence. Overall, this balance between model complexity and predictive accuracy demonstrates the effectiveness of integrating domain knowledge and physical constraints into the neural network-based models. Several specialized techniques have been proposed to further enhance the robustness and stability of training neural ODE type approaches, \eg, stabilized neural ODE~\citep{linot2023stabilized}, random noise injection~\citep{liu2019neural}, Lyapunov loss formulation~\citep{rodriguez2022lyanet}, and multistep penalty neural ODE~\citep{chakraborty2024divide}.

Extending the current framework to accommodate patient-specific geometries with curvature represents an important next step for future research, as it would significantly enhance the clinical applicability of the model. The proposed PCNDE method has already demonstrated its ability to improve the 1D FEM results for idealized geometries. It is reasonable to expect that for more complex, patient-specific geometries, where 1D FEM's accuracy is further reduced~\citep{pfaller2022automated, rubio2024hybrid}, the PCNDE approach could yield even more substantial improvements. \edit{Certainly for such cases the model needs to be retrained with a sufficiently large and diverse training dataset, probably encompassing both idealized and patient specific geometries.} \edit{It is likely that the number of training geometries should be at least on the order of a few hundred in order to achieve good generalizability for patient specific geometries. The inlet boundary condition also needs to be varied, further increasing the number of 3D CFD/FSI cases at least 10-fold. Therefore, such dataset sizes should be at least on the order of 1,000. This poses significant challenges for building such models. First, it requires high computational costs as each patient specific 3D FSI simulation can take tens of hours on multiple CPU cores. Second, current software lack the capability to automatically generate, mesh, and set up models in a fully robust manner without user supervision.} Graph-based approaches have shown promising results for some scenarios~\citep{pegolotti2024learning,sen2024PIGNN}, suggesting a potential avenue for integrating these methods within the PCNDE framework to handle branching network of arteries more effectively. These advancements could pave the way for more accurate and personalized rapid cardiovascular modeling.

Another promising future direction is to develop more sophisticated methods for mapping between 3D and 1D data representations in cardiovascular flows. While cross-sectional averaging has been the traditional approach to reduce 3D data to 1D profiles, it inherently loses critical information about complex flow patterns and spatial variations, which are particularly important in cardiovascular diseases, such as recirculation, vortex formation, and subsequent circumferential variability in wall shear stress. Machine learning-based alternatives, such as autoencoders or neural implicit representations~\citep{sitzmann2020implicit}, could potentially capture and preserve this intricate information more effectively, offering a more detailed hemodynamics model. Future studies should also focus more on uncertainty quantification. An introductory analysis was done for the effect of noise in the input variables in Sec.~\ref{sec:sensitivity_Qin}, but this becomes even more relevant for patient-specific geometries. In these cases, the geometry itself carries underlying uncertainty due to limitations in imaging and segmentation processes. \edit{Cardiovascular geometries exhibit significant variability due to factors like genetics, gender, age, and disease progression. Incorporating probabilistic modeling frameworks could significantly enhance uncertainty quantification by characterizing variability in input parameters, such as imaging-derived vessel geometries or boundary conditions. Rather than providing a single deterministic output, probabilistic models can generate a range of possible outcomes with associated confidence intervals, offering more nuanced and reliable predictions. This capability is particularly valuable in clinical decision-making, where understanding the degree of certainty in predictions can inform treatment strategies. Moreover, stochastic methods could also help generate diverse synthetic datasets, addressing challenges in training data availability.} 
The flexibility of our PCNDE framework opens up exciting possibilities for integrating experimental data, such as 4D Flow Magnetic Resonance Imaging (4D Flow MRI), through additional terms in the loss function. This integration of diverse data sources could lead to more comprehensive and accurate models. Finally, differentiable programming techniques could revolutionize various other aspects of cardiovascular medicine beyond simple hemodynamics modeling. For example, these methods offer the potential to improve medical device design or create more accurate control systems for circulatory support devices. By enabling end-to-end gradient-based optimization, differentiable programming could facilitate more precise parameter estimation, optimization, and control in complex physiological systems, leading to more effective and personalized treatment strategies for cardiovascular diseases.

\section{Conclusions}
\label{sec:conclusion}
In this study, we introduced a novel physics-constrained machine learning technique to create improved one-dimensional models for cardiovascular flows. Our approach, the physics-constrained coupled neural differential equation framework, demonstrated superior accuracy in predicting flow dynamics compared to traditional FEM-based 1D models across a range of inlet boundary conditions and stenosis blockage ratios. A key innovation of our method lies in the spatial formulation of the governing equation for momentum conservation, departing from the conventional temporal approach. This reformulation capitalizes on the inherent temporal periodicity of blood flows, offering a fresh perspective on cardiovascular flow modeling.
These results demonstrate the potential of physics-informed machine learning techniques in advancing cardiovascular flow modeling. The improved accuracy and computational efficiency of our approach open up new possibilities for rapid, patient-specific simulations that could be invaluable in clinical settings. These promising outcomes not only advance hemodynamics modeling but also encourage further research in related fields. Notably, our approach to learning stable systems of coupled PDEs with time-varying boundary conditions has broad applicability beyond cardiovascular modeling. Its generic nature and low computational cost of training suggest potential use across a wider spectrum of time-periodic transport problems in various scientific and engineering domains.

\section{Appendix: Sensitivity to Undeformed Area}
\label{sec:sensitivity_Su}
\begin{figure}[h]
    \centering
    \includegraphics[width=\linewidth]{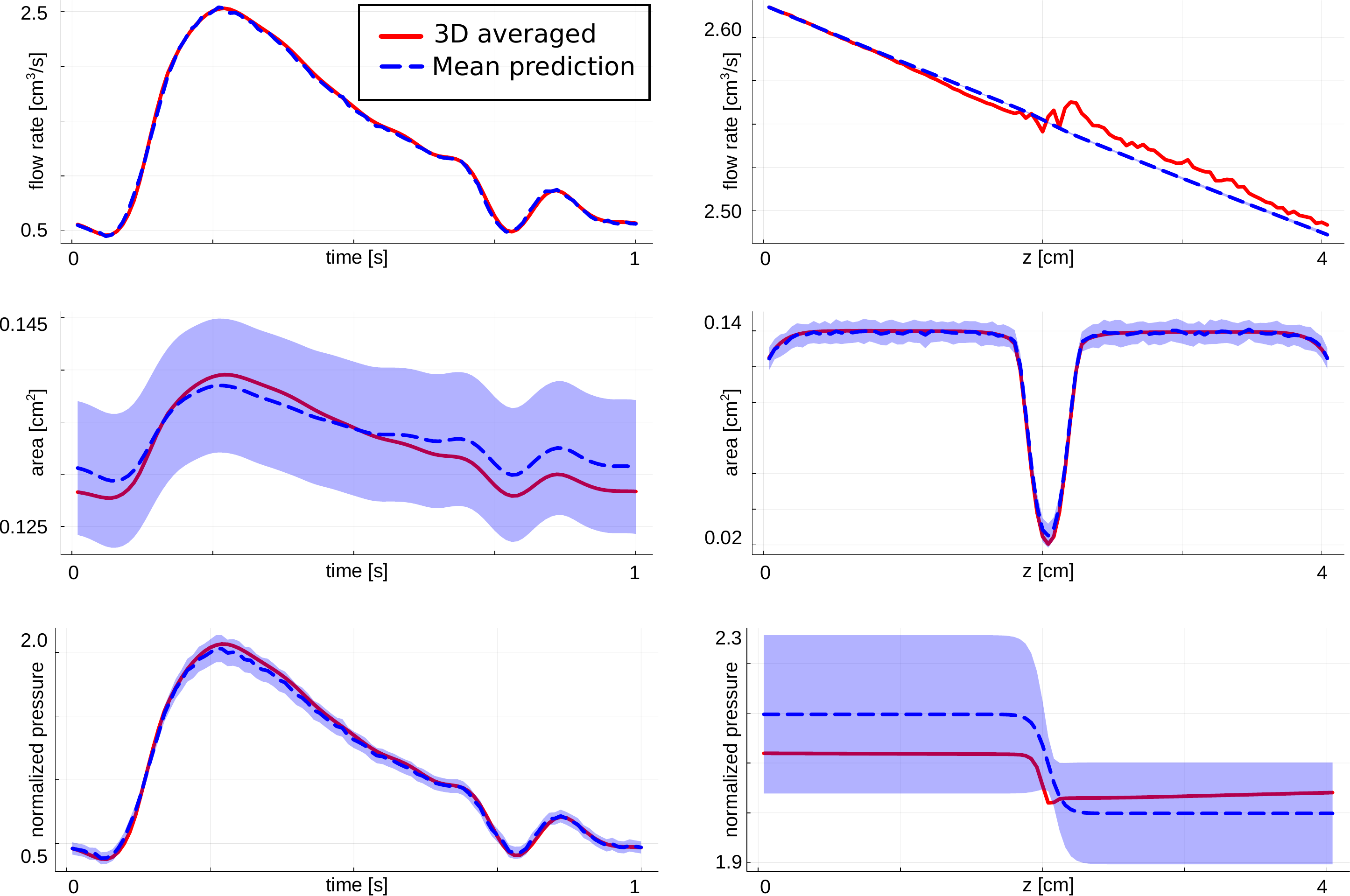}
    \caption{Uncertainty visualization given a noisy undeformed area \(S_{u}(z)\) is shown for flow rate, area, and normalized pressure results. The left column shows the temporal plots at \(z=3.6\) cm, while the right column the spatial plots at \(t=0.25\) s. The dashed blue lines represent the mean of the predictions, while the shaded regions correspond to the standard deviation of the predictions. }
    \label{fig:uncertainty_Su}
\end{figure}

To investigate the effect of noise in the undeformed cross-sectional area, a further sensitivity study was performed. Similar to the inlet flow rate waveform, random noise from a standard normal distribution was added to the undeformed area, scaled by \(\sigma = 0.05 \max(S_{u}(z))\). The corresponding results are shown in Fig.~\ref{fig:uncertainty_Su}. The area results show larger uncertainty than in the case of noise in the inlet flow rate, but the flow rate and pressure panels show lower uncertainty. The average standard deviation of the area predictions relative to the maximum area was \(0.05\), which corresponds to the added noise. For flow rate, the uncertainty is very low, suggesting that the model \(f_{\theta}(Q,S)\) mostly relies on the flow rate and only weakly on the area values. The average standard deviation of the flow rate prediction normalized by the maximum flow rate is \(0.00014\). The uncertainties in the pressure model are larger than the flow rate, but still relatively small. The mean standard deviation of the pressure predictions normalized by the maximum pressure is \edit{\(0.033\)}. However, the mean pressure prediction overestimates the pressure drop at the stenosis. This suggest that the area information is more important to capture the correct pressure drop, while the flow rate information plays a larger role in capturing the temporal evolution of pressure.

\edit{
\section*{CRediT authorship contribution statement}
\textbf{Hunor Csala:} Conceptualization, Data curation, Writing – original draft, Writing – review and editing, Methodology, Software. \textbf{Arvind Mohan:} Conceptualization, Writing – review and editing, Methodology, Supervision. \textbf{Daniel Livescu:} Writing – review and editing, Methodology, Supervision, Funding acquisition. \textbf{Amirhossein Arzani:} Conceptualization, Writing – original draft, Writing – review and editing, Methodology, Supervision, Funding acquisition.

}

\section*{Data Availability}
\edit{The Julia source codes used to generate the results in the manuscript are available on GitHub: \url{https://github.com/amir-cardiolab/PCNDE}. } 

\section*{Conflict of Interest}
The authors have no conflicts. 

\section*{Acknowledgment}
The authors acknowledge funding from the Los Alamos National Laboratory LDRD program office and the National Science Foundation (NSF) award \#2247173. \edit{This work is approved for public release by the Los Alamos National Laboratory under LA-UR-24-32004.}



\cleardoublepage
\phantomsection
\bibliographystyle{elsarticle-num-names}
\bibliography{mybib.bib}


\end{document}